\documentclass{aa}   
\usepackage{epsfig,psfig,natbib,txfonts}
\usepackage{graphicx}
\usepackage{txfonts}
\newcommand{\magarc}{mag. arcsec$^{-2}$\,}

\begin{document}
   \title{Truncation of stellar disks in galaxies at z~$\approx$~1
   \thanks {Observations carried out using the Very Large Telescope at the ESO Paranal Observatory, Program ID 170.A-0788 }}

   \author{I. P\'erez  \inst{1} }

   \offprints{I. P\'erez Mart\'{\i}n \\ email: isa@astro.rug.nl}

   \institute{Kapteyn Astronomical Institute, University of Groningen,
   Postbus 800, 9700AV Groningen, the Netherlands}

   \date{}

   \abstract{ We report here the first evidence for stellar disk
   truncation at high redshift, based on surface photometry of a
   sample of 16 high redshift (0.6~$<$~z$~<$~1.0) disk galaxies from
   the GOODS HST/ACS data. The radial profiles are best fit by a
   double exponential profile. This result agrees with the profile of disks
   in local galaxies. The cosmological surface brightness dimming at
   this redshift range only allows us to detect galaxies with
   spatially ``early'' truncation, R$_{\rm br}$/h$_{\rm in}\leqq$
   3.5. Six galaxies show the radial double exponential structure,
   with an average value of R$_{\rm br}$/h$_{\rm in}\approx$~1.8. Such
   ``early'' truncated galaxies are missing in local samples so
   far. This result opens the ground for observing directly disk
   evolution through the study of the truncation radius as a function
   of redshift.   }

   \maketitle
%

\section{Introduction}
The finite size of stellar disks was first reported by~\cite{kruit79}
who showed that disks were sharply truncated at around 4
scale-lengths. Recently, evidence has arisen that this truncation does
not occur sharply but instead radial profiles are best fit by a double
exponential shape \citep{degrijs01,pohlen02}. For nearby objects, the
profiles obtained from a sample of edge-on galaxies show a two-slope
exponential with R$_{\rm br}$/h$_{\rm in}$ in the range
$3-5$~\citep{kregel}, where R$_{\rm br}$, the break radius, is the
radius where the slope changes and h$_{\rm in}$ is the inner disk
scale-length (see Pohlen et al., 2004; for a recent review on the
truncation radius). This truncation might be related to a surface
density threshold for star formation~\citep{kennicutt} or might be
related to the maximum specific angular momentum of the sphere from
which the disk collapsed~\citep{kruit87}. While galaxy formation
models have been able to reproduce the observed double exponential
profile~\citep{bosch01,zhang}, none of these models have considered
evolution of the truncation radius with redshift. At z~$\approx$~1,
galaxies are half the age of local objects and one can probe into a
period where evolutionary effects may be significant ~\citep{mao}. A
study of the evolution of the truncation radius can help to discern
between truncation arising as a relic from galaxy formation or due to
a star formation threshold. Such a project has been carried out as a
pilot survey using HST archival data and is reported in this Letter.
\section{ Sample and profile fitting}

The cosmological dimming ($\propto$(1+$z$)$^{4}$) determined our
choice of the maximum redshift to reach the outermost isophotes
($\mu_{B}\approx$~25.5\magarc). The  lower redshift limit was chosen
to ensure a reasonable number of objects in the final sample, at a
redshift where evolutionary effects should still be
significant~\cite{mao}. A range of 0.6~$<$~z~$<$~1.0 was finally
chosen. This range implies a dimming between $2-3$ magnitudes with an
isophotal threshold range between $25.5-24.5$~\magarc. Our sample
consists of 16 galaxies selected from the CDF-S and the HDF-N from the
V1.0 GOODS HST/ACS Data Release~\citep{giavalisco}. The HST/ACS $z$-band images were used
for the analysis to ensure that the rest frame of the above galaxies
($V$ and $B$-band) corresponds to the wavelength of widely observed
local samples.  The redshifts for the southern fields were obtained
from the ESO/VLT GOODS~\citep{vanzella} and the COMBO 17
projects~\citep{wolf} and for the northern fields, from  Keck DEIMOS
spectra ~\citep{wirth}. The galaxies were visually selected to be
relatively symmetric and not strongly interacting. Hubble types range
from Sa to Sc (our visual classification of the HST/ACS $z$-band
images) and galaxy inclination angles (derived from the ellipticity of
the outermost isophote) are all less than 35$^{\circ}$. It should be
emphasised that the sample was not created with the intention of being
used as a complete statistical sample. It is, in fact, an initial
sample to determine whether we can find stellar disk truncation at
high-z and whether the derived values fall within the local range.

Ellipse fitting to the light distribution was performed using the
ELLIPSE task within IRAF.  All the galaxies showed point-like nuclear
regions. So, the center was fixed using the coordinates obtained by
fitting a Gaussian to the nucleus. The position angle (PA) and the
ellipticity ($e$) were left as free parameters in the
fitting. Profiles with fixed PA and $e$ were also obtained, no
influence in the break radius was found. The free PA$-e$ fitting
profiles were chosen to determine whether there were any systematic
relation between the position of morphological features with the
truncation radius. No attempt was made to do a bulge-disk
decomposition, and the disk was assumed to dominate the light profile
beyond~$\approx$~1~scale-length, a good assumption for Sb and later
types ~\citep{dejong}; implications of this fitting procedure are
discussed in Section~\ref{results}. The zero point to convert the
count rate into an AB magnitude for the GOODS $z$-band was taken from
the ACS web page (F850LP = 24.84). The sky background was subtracted
by taking the mean value of boxes placed around the galaxies in
regions far from the target galaxies. The standard deviation of the
median of the distribution in the different boxes was adopted as the
error in the background determination.

We then fit two exponential functions of the form $I({\rm R})^{\rm
in/out}~=~I^{\rm in/out}_{0}$~exp$(-{\rm R/h}_{\rm in/out})$ to the
1-D azimuthally averaged profile ($in$ refers to the inner disk and
$out$ to the outer exponential disk). The criterion used for detection
of truncation was that the outer exponential should extend to at least
2 $\times$ h$_{out}$ before reaching the noise level. The break radius
was first visually determined, with an error around 12\%. This value
was used to separate the fitting regions for the inner and the outer
exponentials; excluding the bulge region. The final break
radius is defined as the radius where the two fitted exponential
profiles cross. Fig.1 shows the fitted functions plotted on top of the
derived profiles. While the error in the inner scale-length is
dominated by the effect of the bulge, the outer scale-length error is
dominated by the determination of the break radius and the
sky-subtraction. Table 1 shows the errors in the outer scale-length
determination when doing the fitting to profiles
$\pm$~1~$\sigma$ of the sky-level including the error in the break radius
determination.

A correction for cosmological dimming was applied (H$_{0}$~=~71
km~s$^{-1}$~Mpc$^{-1}$, $\Omega_{M}$~=~0.3 and
$\Omega_{\Lambda}$~=~0.7 throughout this work including converting the
measured angular sizes to physical sizes using the given redshifts),
with k-corrections  estimated using the results
of~\cite{fukugita}. First, the k-correction for the HST/ACS $z$-band
was computed for galaxy morphological type Sbc for each of the
redshift values, the error in the K-correction is $\pm$0.25~mag. Then,
the $z$-band was transformed to $B$-band assuming the rest-frame
($B$~$-$~$z$)~=~2.14$\pm$0.15 ~\citep{fukugita} for the same
morphological type.

\section {Results and comparison to nearby objects}
\label{results}

Of the 16 galaxies, six show a two-slope profile that we identify with
a truncated disk. This result agrees with that found for nearby
objects~\citep[and references therein]{pohlen01}. The list of galaxies
and the 1-D exponential fitting parameters and errors are presented in
Table 1. The azimuthally averaged (360$^{\circ}$) profiles are shown
in Fig. 1. Averaged profiles of different image segments were also
computed (90$^{\circ}$ and 180$^{\circ}$ sections) to check for
systematic differences but no significant changes were found in the
profiles. The R$_{\rm br}$/h$_{\rm in}$ occurs at 1.8$\pm$ 0.5.
Unsharp-masking was used to locate the position of the end of the
spiral structure and to compare it to the R$_{\rm br}$ radius. In two
cases (123637+621159 and 033251$-$27504) the truncation seems to be
related to the end of the spiral arms. The surface brightness at which
the break occurs happens at a mean value of~21.7~\magarc (rest-frame $B$-band).

To compare the value obtained for the high-z galaxies (R$_{\rm
br}$/h$_{\rm in}$ ranging between 1.3-2.2) with the value for nearby
objects we used two different local samples. The mean value of R$_{\rm
br}$/h$_{\rm in}$ obtained for a sample of three nearby face-on
galaxies~\citep{pohlen02} is 3.9$\pm$~0.7. Another result from the
analysis is that the h$_{\rm in}$/h$_{\rm out}$ ratio obtained for the
high-z galaxies (2 $\pm$ 0.9) is remarkably similar (2 $\pm$ 0.2) to
that obtained by~\cite{pohlen02}. Notice the small average R$_{\rm
br}$/h$_{\rm in}$ obtained for the high-z galaxies compared to the
value for local galaxies; this will be referred as ``early''
truncation throughout the paper. It is interesting to compare to this
small sample because of the similarities to our sample in the
methodology used and the galaxy characteristics. A larger sample of
nearby edge-on galaxies (20 galaxies) was studied by~\cite{kregel}
with a mean value of R$_{\rm br}$/h$_{\rm in}$~=~4.0$\pm$1.1. To make
a fair comparison we have to convert these values, obtained in the
$I$-band, to the $B$-band ~\citep{degrijs}. The ratio h$_{\rm
in,B}$/h$_{\rm in,I}$ has been estimated to be 1.32~$\pm$~0.24 for all
morphological types~\citep{degrijs}, taking into account this value we
can re-calculate the local value for the $B$-band, giving a average
R$_{\rm br}$/h$_{\rm in}$~=~2.9$\pm$0.9. In this way the highest
values found for the high-z galaxies would fall in the low-value tail
of the local galaxies; however, the lowest values still fall below the
local distribution.  One should also point out that the methodology
used by~\cite{kregel} to extract the different parameters is different
to the one adopted in this work. Another point to notice is that there
is still no clear result about the dependence of the break radius with
wavelength in local galaxies.

The remaining ten galaxies show no truncation down to R$_{\rm
br}$~=~$3.5-4$~h~(depending on the galaxy). The derived scale-lengths
and central surface brightnesses from the one-exponential fitting are
comparable to those of the ``early'' truncated galaxies. These
galaxies either do not show truncation or might be the counterparts of
local galaxies, with spatially ``late'' truncation. 

The central disk surface brightness derived for all 16 galaxies of the
sample is a few magnitudes brighter than the Freeman value
21.6~-~$\overline{\mu}_{0,B~{\rm rest-frame}}$~=~2.0~$\pm$~1.0 mag
\citep{freeman}. No bulge-disk decomposition was carried out, this
can lead us to overestimate the obtained central surface
brightness. From a test performed on a sample of local galaxies
ranging from Sa to Sc we estimated the maximum error from not taking
into account the bulge to be $\approx0.2$ mag. The errors in our central
surface brightness estimates are high due to the k-correction and the
bulge-disk decomposition errors. However, correcting for this still
cannot account for the difference of around 2 magnitudes found for the
central surface brightness.

As an internal check, we determined the same parameters for the three
face-on galaxies of ~\cite{pohlen02} using their images and a
comparable strategy to the one used for the high-z galaxies. We
obtained similar parameters for the break radius and the scale-lengths
as the ones derived by ~\cite{pohlen02}, although there are some
differences in the techniques used.
\begin{table*}
\begin{center}
\caption{Galaxy parameters$^{1}$}
\label{tab:galaxies}
\begin{tabular}{lllllllllllll}
\hline\hline 
ID& z &h$_{\rm in}$& h$_{\rm in}$& error& h$_{\rm out}$&error& R$_{\rm br}$ &error& $\mu^{\rm in}_{0,B_{\rm~restframe}}$ &$\mu^{\rm br}_{B_{\rm~restframe}}$&R$_{\rm br}$/h$_{\rm in}$& error \\ 
&&(kpc)&(arcsec)&(arcsec)&(arsec)&(arcsec)&(arcsec)&(arcsec)&&&&\\
\hline 
123610+621334&0.69&6.1&0.59&$\pm$~0.14&0.33&$\pm$~0.10&0.99&$\pm$~0.18&19.7&21.7&{\bf 1.7}&$\pm$~0.5 \\
123637+621159&0.78&11.0&0.99&$\pm$~0.13&0.36&$\pm$~0.15&1.32&$\pm$~0.08&20.6&21.9&{\bf 1.3}&$\pm$~0.2\\
033251-275044&0.98&3.5&0.28&$\pm$~0.04&0.15&$\pm$~0.07&0.51&$\pm$~0.07&18.9&20.7&{\bf 1.8}&$\pm$~0.3\\
033233-274410&0.67&4.6&0.45&$\pm$~0.2&0.24&$\pm$~0.09&0.97&$\pm$~0.05&19.6&21.8&{\bf 2.2}&$\pm$~1.0\\
123709+622006&1.01&4.5&0.36&$\pm$~0.06&0.18&$\pm$~0.08&0.48&$\pm$~0.08&19.2&20.5&{\bf 1.3}&$\pm$~0.3\\
123708+621252&0.84&5.9&0.51&$\pm$~0.03&0.25&$\pm$~0.13&1.10&$\pm$~0.16&19.6&21.7&{\bf 2.2}&$\pm$~0.3\\
\hline
\end{tabular}
\end{center}
\end{table*}
\section {Discussion and summary}

High-z galaxies are dustier than local galaxies. This will influence
the h$_{\rm in}$ making it appear larger. The true h$_{\rm in}$ would
need to be 50\% smaller in order to raise the minimum R$_{\rm
br}$/h$_{\rm in}$ for the high-z galaxies to the lowest values seen
locally, R$_{\rm br}$/h$_{\rm in}\approx$~2.6~\citep{kregel}
($\approx$~2~in the $B$-band). In the nearby universe, the
scale-lengths show a systematic increase at shorter wave-lengths but
not steeply~\citep{beckman}. It is not clear what is the dust
contribution and distribution in galaxies at high-z which prevent us
at this point to predict which would be the expected R$_{\rm
br}$/h$_{\rm in}$ for the high-z galaxies. It is interesting to notice
that the ratio between the inner and the outer scale-length seems to
remain unchanged for the high-z sample which suggests that the dust
does not make R$_{\rm br}$/h$_{\rm in}$ appear smaller at higher
redshifts.

In a recent paper,~\cite{pohlen04} report a connection between ``early''
truncation and bars, where the truncation is associated with the outer
Lindblad resonance (OLR). This implies that the break happens at
R$_{\rm br}$/h$_{\rm in}~\approx$~$1-2$, consistent with the values
found here. However, there is no clear evidence for the presence of
bars in the 6 galaxies of our sample. The isophotes for the bar region
are characterised by a constant PA while the $e$
plot increases, reaching a maximum at the end of the bar
~\citep{wozniak}. We do not observe this behaviour for the sample
galaxies. However, dust might be obscuring the bar in the rest
$B$-band. We could then also estimate, using the relation between the
OLR and the truncation and relating the position of the OLR to the
galaxy morphology, whether the systems are fast or slow rotators. One
should also point out that some clearly barred objects did not show
sign of truncation. This fact is also observed in nearby objects
~\citep{pohlen04} and should be further investigated.

In summary, a double exponential provides a good fit to the profiles,
similar to nearby galaxies. Six of the sixteen high-z galaxies show
truncation with an average R$_{\rm br}$/h$_{\rm in}$~$\approx$~1.8,
while for nearby objects R$_{\rm br}$/h$_{\rm in}$~$\approx$~4,
reflecting several biases; our R$_{\rm br}$/h$_{\rm in}$ detection
limit and the effect of dust. However, there are no local counterparts
to the ``earliest" truncated disks. The existence of these ``early''
truncated galaxies at high-z suggests a critical star formation
density model for the origin of truncation, with disks forming
inside-out. Longer wavelength observations are needed to minimise and
characterise the effects of dust. A larger and deeper survey could
probe disk profiles to R$_{\rm br}>$~4~h$_{\rm in}$ (where the break
radius is observed for nearby objects) to obtain proper statistics on
the frequency and distribution of truncation at high-z and to probe
whether the 10 remaining galaxies are untruncated or whether they show
break radii similar to those of local galaxies.

\begin{acknowledgements}
I am grateful for the useful discussions with M. Pohlen and M. Kregel
and to Pohlen again for kindly making his data available. I am
indebted to N. Kanekar M. Verheijen and R. Peletier for their support
and the careful reading of the manuscript. I also thank the anonymous
referee for constructive comments that helped to greatly improve this
Letter.
\end{acknowledgements}
\bibliography{ref_Gg193}
\bibliographystyle{natbib}

\begin{figure*}
\begin{center}
\vspace{0cm}
\hspace{0cm}\psfig{figure=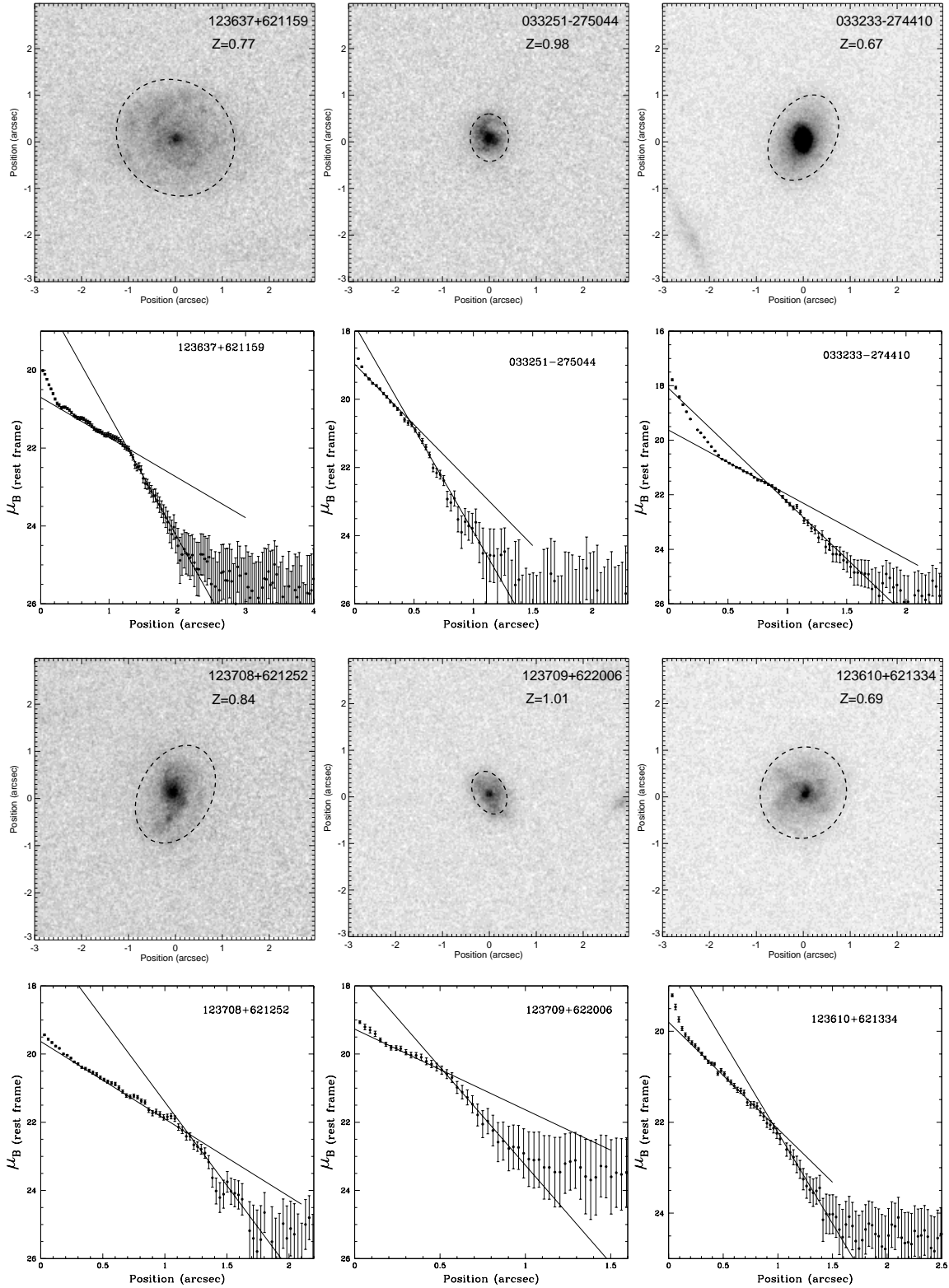,width=17.0cm}
\vspace{-5mm}
\caption[Time evolution of the gas distribution in the NGC 7483
simulation]{Azimuthally averaged surface brightness profiles and
$z$-band images showing the fitted exponential functions (K-correction and cosmological dimming correction applied) for the 6 galaxies that show truncation. The location of the break radius has been overlaid on the $z$-band images.}
\label{fig:ngc7483evolgas}
\end{center}
\end{figure*}

\end{document}